\documentclass[fleqn,10pt]{wlscirep}
\usepackage[english]{babel}
\usepackage[autostyle]{csquotes}
\usepackage{verbatim}
\usepackage{textcomp}

\usepackage{graphicx}
\usepackage{amssymb}
\usepackage{amsthm}
\usepackage{amsmath}
\usepackage{mathtools}
\title{Altered Modularity and Disproportional Integration in Functional Networks are Markers of Abnormal Brain Organization in Schizophrenia}

\author[1,+]{Cinelli, M.}
\author[2,3,+]{Echegoyen, I.}
\author[4,+]{Oliveira, M.}
\author[5,+]{Orellana, S.}
\author[6,*]{Gili, T.}

\affil[1]{University of Rome "Tor Vergata", Rome, Italy}
\affil[2]{Centre for Biomedical Technology, Technical University of Madrid, Madrid, Spain}
\affil[3]{Rey Juan Carlos University, Madrid, Spain}
\affil[4]{GESIS -- Leibniz Institute for the Social Sciences, Cologne, Germany}
\affil[5]{Utrecht University,  Utrecht, Netherlands}
\affil[6]{IMT School for Advanced Studies, Lucca, Italy}
\affil[*]{tommaso.gili@imtlucca.it}

\affil[+]{These authors contributed equally to this work.}


\begin{abstract}
Modularity plays an important role in brain networks' architecture and influences its dynamics and the ability to integrate and segregate different modules of cerebral regions. Alterations in community structure are associated with several clinical disorders, specially schizophrenia, although its time evolution is not clear yet. In the present work, we analyze fMRI functional networks of $65$ healthy subjects (HC) and $44$ patients of schizophrenia (SZ), $28$ of them in a chronic state (CR) of illness, and $16$ at early stage (ES). We find clear differences in edges' weights distribution, networks density, community structure consistency and robustness against edge removal. In comparison to healthy subjects, we found that networks from SZ patients exhibits wider weight distribution, larger overall connectivity, and are more consistent in the community structure across subjects. We also showed that the networks of SZ patients tend to be more robust to edge removal than healthy subjects, while having lower network density. In the case of early stages patients, we found that their networks exhibit topological features consistently in between the ones obtained from the other two groups, resulting in a tendency towards the chronic group state.
\end{abstract}
\begin{document}

\flushbottom
\maketitle

\thispagestyle{empty}

\section*{Introduction}

The structure and function of the brain can be conveniently characterized as a network of interacting neurons, cortical columns, or brain regions, depending on the spatial scale considered\cite{Sporns2011}. 
Both structural and functional brain networks exhibit emerging features commonly encountered in other networked systems, such as small-world structures, hubs, high assortativity, that leads to a rich-club of highly connected hubs, or modules\cite{Bullmore2009,Bassett2017,Buldu08}. 
These properties are important to ensure an efficient adaptation to the environment, and changes on them are associated with diverse pathologies\cite{bassett_human_2009,griffa_structural_2013,kaiser_potential_2013}.

Schizophrenia is a severe psychiatric disorder that manifests through positive and negative symptoms. Positive symptoms are, among others, hallucinations, disorganized speech and delusions; whereas negative symptoms are grossly disorganized or catatonic behavior, diminished emotional expression and anhedonia. The disease is commonly accompanied with alterations in sleep, inappropriate affect, depersonalization or derealization \cite{american_psychiatric_association_diagnostic_2013}, and can have an insidious or sudden onset. 

In classical neuroscience, researchers have developed different theories to explain the processes behind SZ, from changes in neurotransmitters, mainly dopamine, glutamate, serotonin, and $\gamma$-aminobutyric acid (GABA)\cite{SchizoTransmitt}, to the conception of the disease as a breakdown in the integration of brain's functionally specialized processes\cite{fodor1985precis, david1994dysmodularity}. Neuroimaging studies support these theories. Regarding functional integration, the network architecture in SZ patients tend to be more disconnected than in healthy populations\cite{fornito2012schizophrenia, van2014brain}.

More concretely, modularity and network partitioning has proven to play an important role in shaping different topologies between subjects suffering from SZ and healthy population\cite{sun2017modular,reinen2018human}. 
A module refers to a set of nodes (i.e., brain regions) exhibiting higher connectivity among themselves in comparison to nodes outside the set\cite{newman2004finding}. Modules in the brain have been shown to be related to integration of information\cite{Sporns2013}. Healthy brains are consistently  found to be within a dynamic balance between hierarchical integration and segregation of information which depends, among others, on task requirements and level of arousal\cite{Buzsaki2006,Sporns2016}. This implies that information must flow to specific areas, depending on the task, and should be confined inside to be integrated for further functions, not spreading along the whole cortex. Disturbances of such balance are associated with several clinical disorders\cite{bullmore_economy_2012}. 
This relationship occurs in structural and in functional brain networks\cite{meunier_hierarchical_2009,tononi_measure_1994}, though the link between structure and function is far of being unveiled. In terms of modularity, structure and function are highly consistent, as analyses of fMRI networks tend to agree systematically\cite{meunier2010modular,hagmann_mapping_2008,chen2008revealing}.

Still, different types of patients exhibit distinct features in their brain networks. In childhood onset schizophrenia, intra-module connections are lower and extra-module connections are higher than what is expected in healthy population\cite{alexander2010disrupted}. 
In this case, modules tend to be less clearly bounded and the structural pattern tend to dissolve, preventing proper integration of information. The literature, however, diverges on adult onset schizophrenia, with results signaling both altered and intact overall modularity\cite{yu2012modular, lerman2016network}.
Yet, findings on community structure (i.e., communities in which different nodes participate) are consistent regardless of diagnostic status\cite{alexander2012discovery, yu2012modular, lerman2016network}. In these studies, however, a small but not negligible percentage of nodes is systematically placed different in schizophrenia, indicating regular changes in topology. Therefore, it is still unclear how modularity diverges in these clinical populations. 


In such a complicated framework our aim is to unveil the differences concerning the community structure and the overall robustness in terms of interconnectivity of brain areas of functional networks (i.e. fMRI based) associated to healthy subjects and patients of schizophrenia. The first goal is accomplished by comparing different community structures across classes of patients in order to find cross similarities/differences. The second goal is accomplished by looking at the size of the largest connected component after the removal of links among the different brain areas using a thresholding logic which penalizes those with the lowest weights. Our findings provide insights about actual topological differences among healthy and unhealthy subjects in terms of homogeneity of partitions and fragility of the network with not negligible differences also between two classes of SZ patients (i.e. early stage and chronic schizophrenia). Nevertheless, the obtained results constitute also a base for discerning the quality of different methods concerning the study of fMRI (and in general dense) networks.  

\section*{Methods}

\subsection*{Subjects and clinical assessment}

We recruited forty four patients diagnosed with schizophrenia according to the DSM-IV- TR (APA, 2000) criteria (SZ group). All patients were diagnosed by one senior clinical psychiatrist using the structured clinical interview for DSM-IV- TR (SCID-I/P)\cite{First1997}. Other inclusion criteria were: 1) age between 18 and 65 years; 2) at least 8 years of education; 3) no dementia or cognitive deterioration according to the DSM-IV- TR criteria, and a Mini- Mental State Examination (MMSE)\cite{Folstein1975} score higher than 24, consistent with normative data in the Italian population\cite{Measso993}; and 4) suitability for a Magnetic Resonance Imaging (MRI) scan. Exclusion criteria were: 1) a history of alcohol or drug dependence or abuse in the last two years according to the DSM-IV- TR, 2) traumatic head injury, 3) any past or present major medical or neurological illness, 4) any other psychiatric disorder or mental retardation diagnosis and 5) MRI evidence of focal parenchymal abnormalities or cerebrovascular diseases.

All patients were in a phase of stable clinical compensation. Age at onset was
defined as age at first hospitalization or, when possible, age at onset of positive or negative symptoms prior to the first hospitalization. Out of the initial sample of 44 patients, 16 were considered being at early stages of symptoms (the period from the first psychotic symptoms and the diagnosis was shorter than two years), while 28 were considered chronic patients (the history of illness from the diagnosis of illness was longer than five years).

The Positive and Negative Syndrome Scale (PANSS)\cite{Kay1987} was administered to rate severity of psychopathological symptoms. PANSS ratings were obtained on all information available pertaining to the last week of the assessment. Extrapyramidal side effects due to current treatment were assessed by the Simpson-Angus Rating Scale (SARS)\cite{Simpson1970}. The Abnormal Involuntary Movements Scale\cite{Guy1976} was administered to determine whether tardive dyskinesia was present; however, no patient suffered from this disturbance. All patients were receiving stable oral doses of one or more atypical antipsychotic drugs such as risperidone, quetiapine, or olanzapine. Antipsychotic dosages were converted to estimated equivalent dosages of olanzapine by using a standard table\cite{Woods2003}. 

We also recruited 65 healthy controls in the same geographical area, so that all the subjects recruited for the study were homogeneous with regard to ethnicity and cultural background. They were rigorously matched for age, education and gender with the patients diagnosed as having schizophrenia. All HC were screened for any current or past diagnosis of DSM-IV- TR axis I or II disorders using the SCID-I and SCID-II\cite{First1997}. A diagnosis of schizophrenia or any other mental disorder in first-degree relatives was also an exclusion criterion. 

Our local Ethics Committee approved the study protocol. Written informed consent was obtained from all patients after they received a full explanation of the study procedures.

\subsection*{Data acquisition and processing}

MRI data were collected using gradient-echo echo-planar imaging at 3T (Philips Achieva) using a (T2*)-weighted imaging sequence sensitive to blood oxygen level-dependent (BOLD) signal (TR $= 3s$, TE $= 30ms$, matrix $= 80$x$80$, FOV $= 224$x$224$, slice thickness $= 3mm$, flip angle $= 90^\circ$, 50 slices). A thirty-two channel receive-only head coil was used. A high-resolution T1 weighted whole-brain structural scan was also acquired ($1$x$1$x$1mm$ voxels). Subjects were asked to lay at rest in the scanner with eyes open. For the purposes of accounting for physiological variance in the time-series data, cardiac and respiratory cycles were recorded using the scanner’s built-in photoplethysmograph and a pneumatic chest belt, respectively.

The human brain was segmented into 278 macro-regions from the template proposed by Shen et al.(2013)\cite{Shen2013}. Resting state fMRI signals were averaged across each region to generate 278 time-series. Several sources of physiological variance were removed from each  individual subject’s time series fMRI data. For each subject, physiological noise correction consisted of removal of time-locked cardiac and respiratory artifacts (two cardiac harmonics and two respiratory harmonics plus four interaction terms), using linear regression\cite{Glover2000}, and of low-frequency respiratory and heart rate effects\cite{Birn2006,Chang2009,Shmueli2007}.

fMRI data were then preprocessed as follows: correction for head motion and slice-timing and removal of non-brain voxels\footnote{Removal performed using FSL: FMRIB’s Software Library,  \url{www.fmrib.ox.ac.uk/fsl}}. Using custom software written in Matlab (The Math Work); data were demeaned and detrended, the six parameters obtained by motion realignment and their derivatives were regressed out. Further nuisance regressors were removed as sources of unwanted signals coming from white matter (WM) and cerebral spine fluid (CSF). Specifically, we calculated the first six principal components of the BOLD signal in two eroded masks that included both the ventricles and the white matter region, respectively\cite{Edalat2012}. Head motion estimation parameters were used to derive the frame-wise displacement (FD) and the root mean square value of the differentiated BOLD time series (DVARS): time points with high FD and DVARS (FD $>$ 0.2 mm and DV $>$ 20) were replaced through a least-squares spectral decomposition as described in Power et al. (2014)\cite{Power2014}. Data were then band-pass filtered in the frequency range $0.01-0.1$ Hz, using a zero-phase second order Butterworth filter.

Since the template proposed by Shen et al.\cite{Shen2013} was in the MNI standard space, a two-step registration process was performed. FMRI data were transformed, first, from functional space to individual subjects’ structural space using FLIRT (FMRIB’s Linear Registration Tool) and then non-linearly to a standard space (Montreal Neurological Institute MNI152 standard map) using Advanced Normalization Tools\footnote{ANTs; Penn Image Computing \& Science Lab, \url{http://www.picsl.upenn.edu/ANTS/}}. Finally data were spatially smoothed ($5$x$5$x$5$ mm full-width half-maximum Gaussian kernel).

\subsection*{Functional Networks}

\begin{figure}[t!]
\centering
\includegraphics[width=\columnwidth]{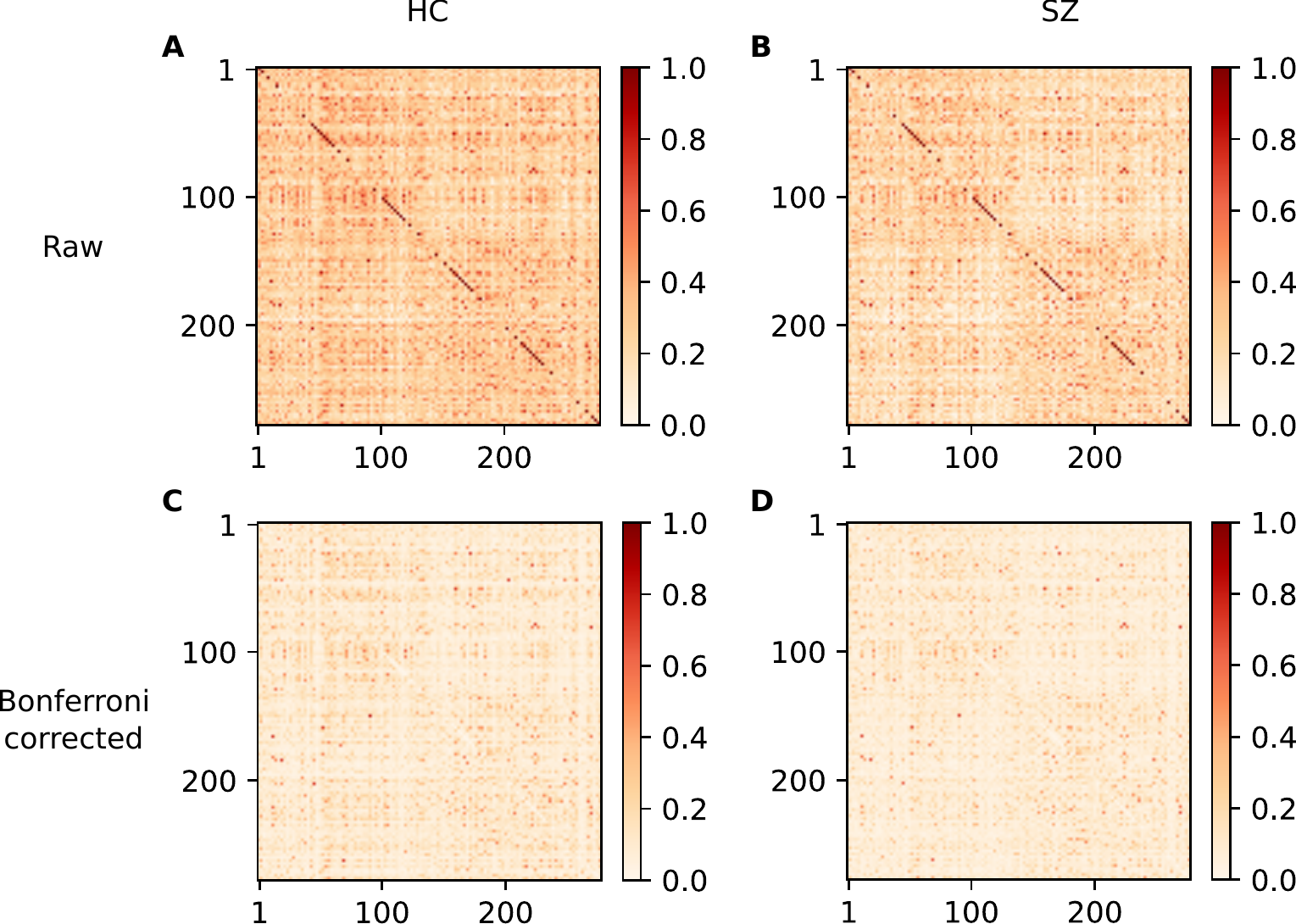}
\caption{\textbf{Average connectivity for HC and SZ.} Upper row matrices correspond to non corrected connectivity. Bottom row matrices contain connectivity after Bonferroni correction. Leftmost matrices correspond to the average connectivity of healthy patients (HC), and rightmost matrices correspond to schizophrenia patients (SZ)}
\label{fig:WeightsMats}
\end{figure}

From each subject we estimate connectivity among regions of interest (ROIs, as extracted from the atlas) pair-wise with Pearson correlation coefficient ($r$). It is known that $r$ is not always the best option, as brain time series are considered non-stationary and non-linearly coupled\cite{Stam2005}. However, at small time scales, stationary and linear approximations to coupling are good enough\cite{Sakkalis2011}. This is precisely the case with the analyzed dataset. Each of the 278 areas considered comprises a time series of 180 samples at a resolution of 3 Hz. For and each subject, we obtain a functional network in which nodes are ROIs and edges are the normalized value of $r$. Normalization is calculated by permutation test, a common procedure of surrogate analysis,  extensively used in neuroscience since the seminal work of Schreiber and Schmitz \cite{Schreiber2000}, and proven to offer a good estimation of the real correlation between time series\cite{Cohen2014,maris2007nonparametric,Nichols2001NonparametricExamples}. The method is as follows: randomize the original time series, compute $r$ and repeat the whole process. $1000$ iterations allow a robust recovery of the distribution of values under the null hypothesis of no relationship between nodes, and approaches a Gaussian distribution as the number of iterations increases. Then, the actual value obtained without randomization ($r_{ij}$) is linearly normalized as a common \textit{z}-score, that is, $\hat{r}_{ij} = \frac{r_{ij} - \mu_{0}}{\sigma_{0}}$, where $\mu_{0}$ is the mean of the null distribution, and $\sigma_{0}$ its deviation; and its probability is calculated. Randomization implies destroying temporal coupling among ROIs while maintaining the distribution of values. Thus, the null distribution obtained randomizing defines the probability of finding a concrete value of the edge by chance, in the case of no temporal correlation. This procedure allows us to set a threshold at $p = 0.01$ and keep only those edges more unlike than this probability. As the number of comparisons increases with the number of nodes ($N$x$(N-1)$), we correct the \textit{p-value} with the Bonferroni correction procedure. Although it is the most conservative method, it is also true that the networks become notably sparser, making community detection much more effective, as only significant edges will be conserved (thus revealing the fundamental structure of the modules). The amount of information loss in comparison with the gain in the ability to detect the communities makes it the best method to correct the p value. The corrected p value is $3.25$x$10^{-7})$. Every edge whose probability is above this threshold is set to zero, and the whole matrix is squared, making positive all edges. Fig. \ref{fig:WeightsMats} shows the result of this procedure for groups averaged connectivity. 

\subsection*{Community Detection and Partitions Comparison}

The process of discovery of naturally emerging groups from the network topology is commonly referred to as community detection. An established way of splitting the network into communities is represented by the modularity ($Q$) maximization\cite{Newman2006}. Modularity is a quantity that represents the correlation among the node attributes (i.e. assignment to communities) computed across network edges; it can be considered as a proxy for the quality of partitions. Therefore, modularity represents a quantity that needs to be optimized when performing community detection and it can be written as:
\begin{equation}
Q = \sum_{i,j} (A_{i,j} - \frac{k_ik_j}{2m})\delta(c_i,c_j)
\end{equation}
where $A_{i,j}$ is the adjacency matrix (or weights matrix) of the network, $k_i$ is the degree (strength) of the node $i$, $c_i$ represents the community to which vertex $i$ is assigned, $\delta$ is the Kronecker delta function and $m$ is the number of network links.

Among the various algorithms for modularity maximization, we adopt an heuristic method called the Louvain method \cite{blondel2008fast}. Usually, it provides high results in terms of modularity, thus high quality of the discovered community structure and fast computation times. The Louvain approach, however, may present suffers from the resolution limit, similarly to all the other methods which employ modularity maximization\cite{fortunato2007resolution}. The resolution limit determines the performance in the identification of relatively small communities, and may yield poor results in some cases. Thus, in our study, we also detect communities using the so-called Surprise algorithm\cite{aldecoa2013surprise}, enabling us to evaluate how the resolution limit problem affects the quality of the retrieved community structure.

Surprise algorithm works by maximizing a quantity called Surprise that represents the probability to find edges between nodes in the same partition (i.e., community). This measure can be defined using an information-theoretic perspective with the Kullback-Leibler divergence, allowing a fast computation\cite{traag2015detecting}. The algorithm maximizes this measure by modifying the partition assignment in the network and aggregating the communities. This procedure overcomes the resolution limit issue\cite{traag2015detecting} and gives us meaningful partitions of the processed topologies. 

In order to compare the various partitions in communities (one per subject) we use the rand index \cite{rand1971objective}. This  measure gives us the similarity among clusters based on the frequency of elements sharing the same/different allocation. We exploit the rand index instead of its adjusted version since we are comparing partitions obtained with the same method across different networks (and not different methods of community detection on the same network).

\section*{Results}

Density in networks changed from around 0.735 (normalizing by the number of nodes) to around 0.52 after applying Bonferroni correction. This can be qualitatively observed from the heatmaps displayed in Fig.~\ref{fig:WeightsMats}. 

\begin{figure}[b!]
\centering
\includegraphics[width=\columnwidth]{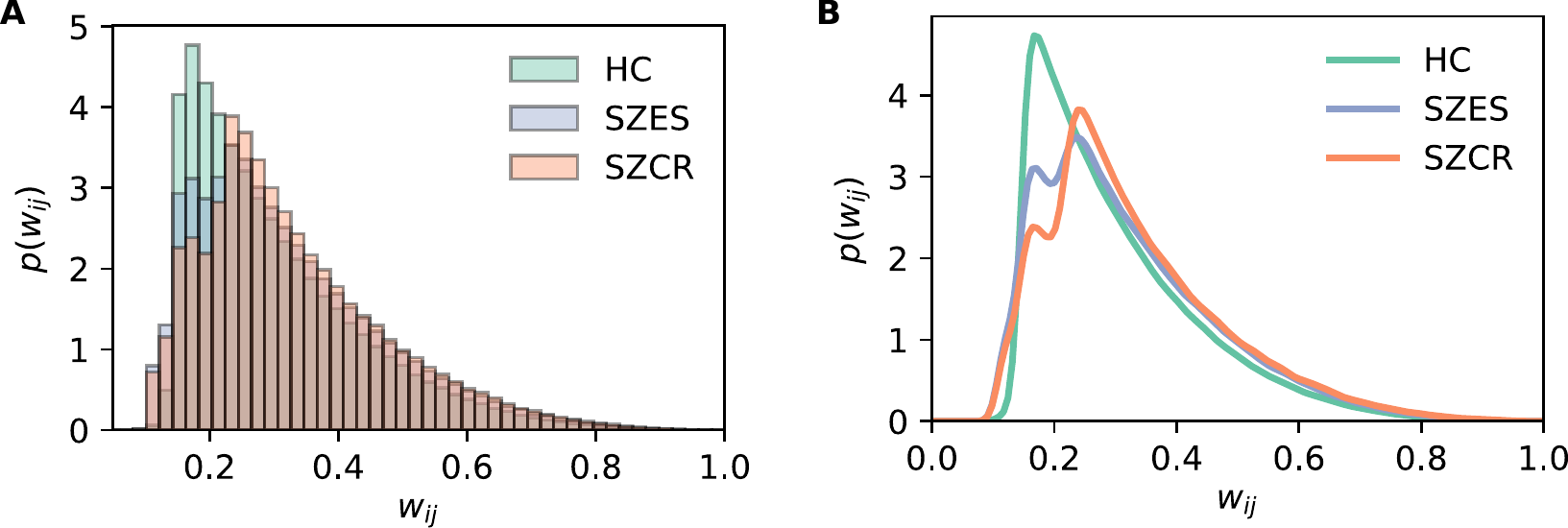}
\caption{\textbf{Distribution of weights for the HC and SZ (ES and CR) groups}. There are significant differences between all groups' distributions (\textit{p $< 0.001$}). SZ (CR) patients exhibit a wider distribution, with higher overall connectivity. ES patients are in between the other two groups.}
\label{fig:PDFsWeights}
\end{figure}

As shown in Fig.~\ref{fig:PDFsWeights}, the distribution of weights is different between groups. Confirming this qualitative observation, a KS test of distribution differences yield a \textit{p value} of 0.0001, indicating to reject the null hypothesis of equal distributions. The same result in hypothesis testing was obtained with a previous shuffling of 200 iterations (p value $= 0.001$) Average level of synchronization is higher, and the distribution is wider. In the HC, values of synchronization are lower, but more concentrated.

Interestingly, differences are clearer when we take into account the two different SZ groups (early stage and chronic state). This is due to the fact that the early stage group has a more bimodal distribution, with clear tendency toward both groups (HC and SZ chronic). As the early stage group is composed of patients that has been suffering the disease and receiving treatment for a shorter period of time, it is reasonable to interpret that as time goes by, the disease stabilizes at a different distribution of weights (a different patterns of synchronization). Hence, what we are observing in the early SZ group is the transition to that point. This increase in synchronization could be interpreted as a compensatory mechanism for the alterations in communities, as other studies have reported\cite{sun2017modular}. It also may be an indirect effect of medication, as neurotransmitters play a major role in populations' coupling.

\begin{figure}[h!]
\centering
\includegraphics[width=5.5in]{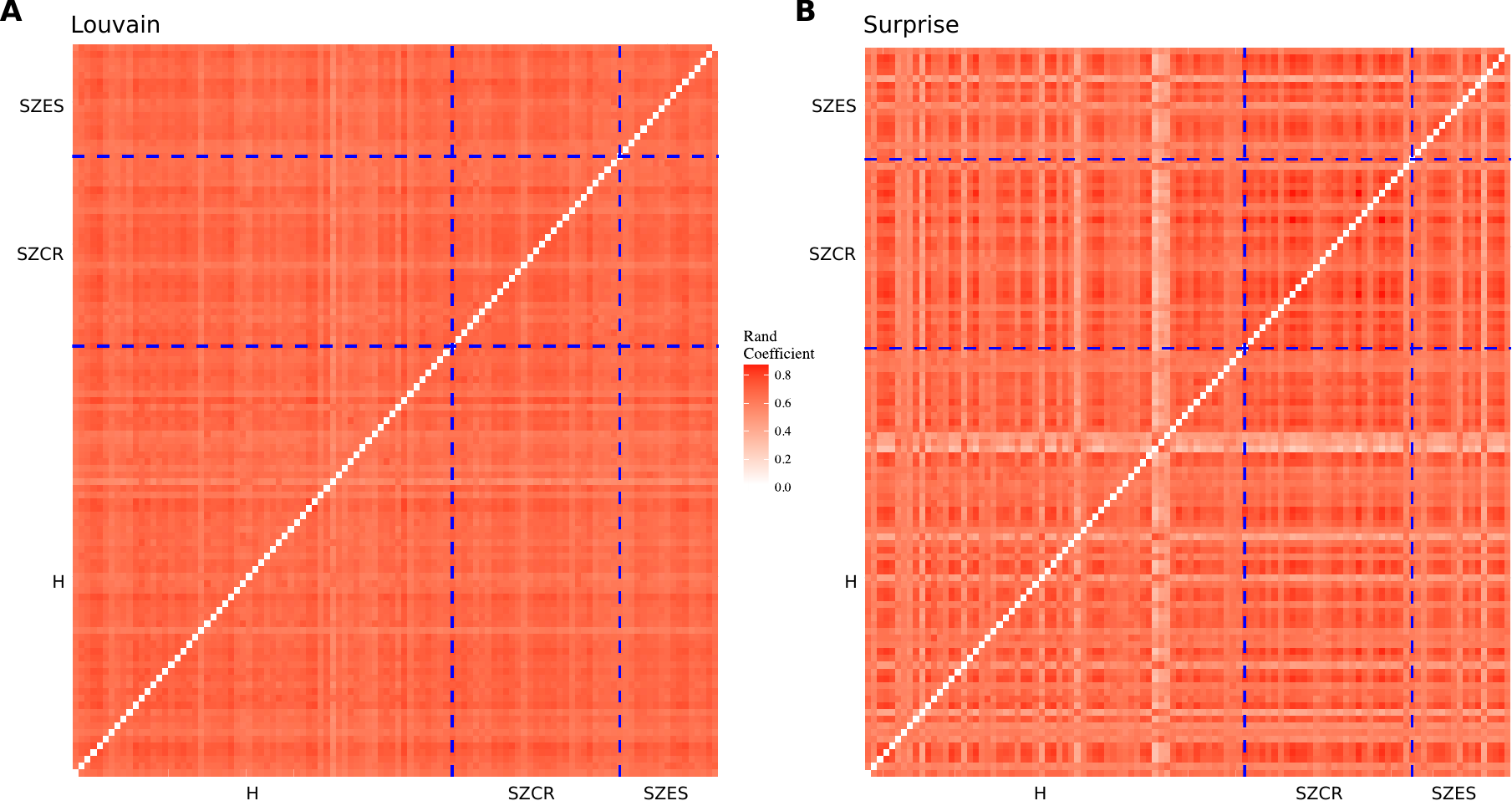}
\caption{\textbf{The rand coefficient related to the community structure for different subjects.} The Louvain algorithm (\textbf{A}) is unable to detect difference between classes of patients, while the Surprise algorithm  (\textbf{B})  identifies homogeneity between individuals of the same class. }
\label{fig:Rand}
\end{figure}

\begin{figure}[h!]
\centering
\includegraphics[width=5.2in]{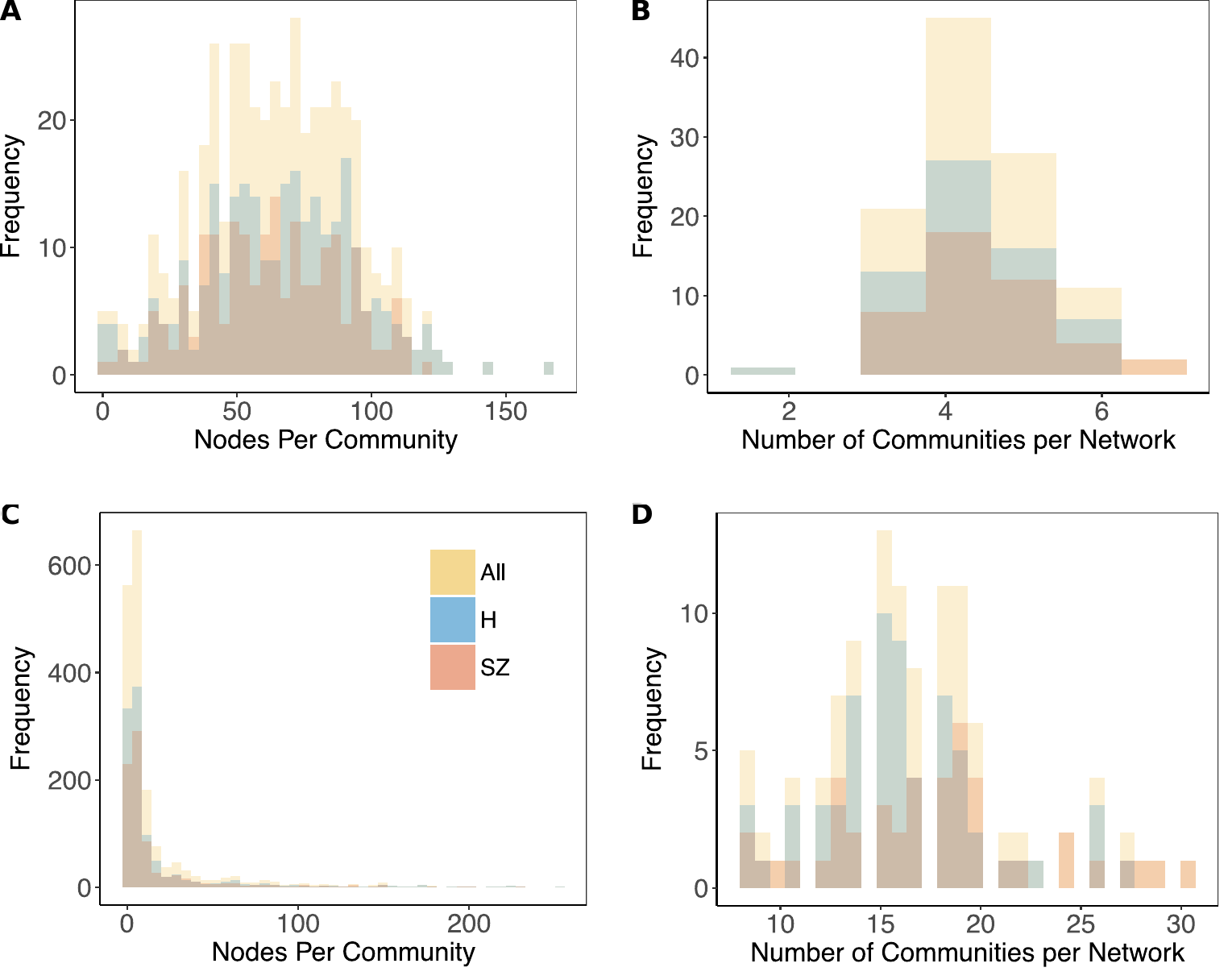}
\caption{\textbf{Composition of communities found in the networks.} The Louvain algorithm (\textbf{A} and \textbf{B}) tends to find smaller communities than the Surprise algorithm (\textbf{C} and \textbf{D}).}
\label{fig:CompositionCommunities}
\end{figure}

Fig.~\ref{fig:Rand} shows the output of the comparison between the different partitions into communities by means of two algorithms, namely Louvain and Surprise. We firstly note how the Louvain method (Fig.~\ref{fig:CompositionCommunities}A and Fig.~\ref{fig:CompositionCommunities}C) suffers from the resolution limit being able to report fewer communities with a higher number of nodes as shown in Fig.~\ref{fig:Rand}. Conversely, partitions obtained with the Surprise method are characterized by a lower number of nodes and are thus able to capture more precisely the inner modular structure of the considered functional networks. The comparison among partitions obtained with the Surprise method suggest a higher similarity among the topologies related to schizophrenic patients shedding light on the relative similarity across such functional networks.

Both community detection methods yield similar results, although the surprise algorithm allow a better understanding on the consistency of communities. SZ patients have more consistent communities, meaning that communities tend to recruit the same nodes, while the HC group shows a higher heterogeneity. This could mean that SZ patients have a more rigid configuration of communities, loosing flexibility; and vice versa, the HC group shows a more flexible configuration. This is indicated by the rand index, a validated method to compare communities among groups. 

\begin{figure}[h!]
\centering
\includegraphics[width=5.2in]{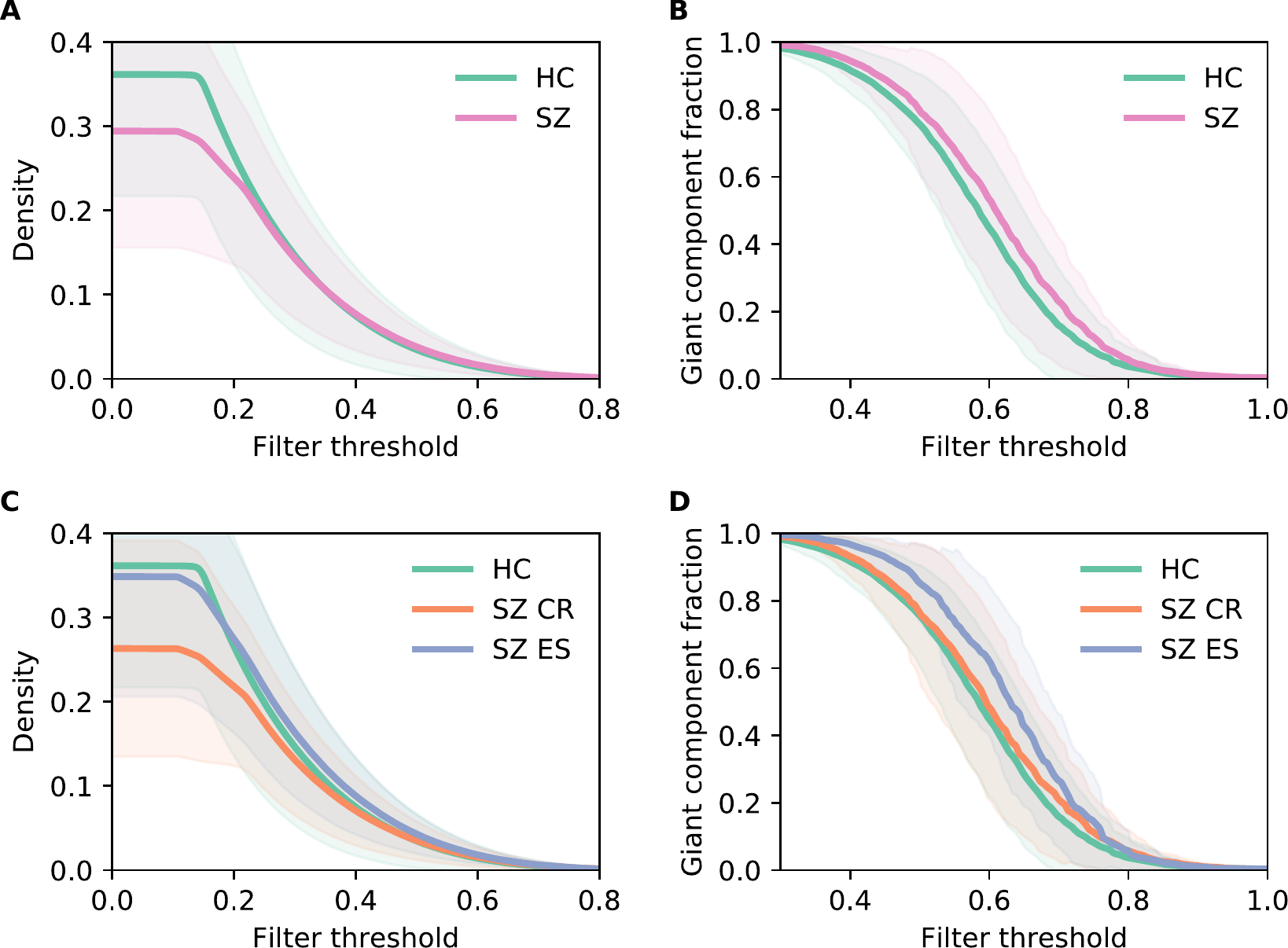}
\caption{\textbf{Effect of the threshold on density and giant component.} Upper row show values for HC and SZ, and bottom row distinguishes also SZ subgroups (CR and ES). Leftmost plots display changes in edge density as a function of the threshold, while rightmost plots do for the size of the giant component}
\label{fig:Robustness}
\end{figure}

Fig.~\ref{fig:Robustness} depicts the effect of raising the threshold of functional connectivity. As the threshold increases, density and size of the giant component decreases. Given that average synchronization is higher in the SZ group, the giant component is more robust than in the HC group. Surprisingly, density is notably smaller in the chronic group in comparison to the other two groups. HC group is more dense, and early stage SZ group is in-between, indicating that, due to the disease or to its medication, density tends to decrease with years, making the giant component more robust. That is, even though functional networks are less dense in SZ patients, they are also more robust to disconnection of the giant component, and this tendency increases with time. As far as we know, this is a novel finding not reported before in the literature. 

\clearpage

\section*{Discussion and Stylized Facts}

In the present paper we have shown major differences in weight distributions between SZ (ES and CR) and HC, and reported a novel finding: while SZ patients functional networks' are less dense than healthy ones, the giant component tend to be more robust to edge removal. Also, communities tend to be more consistent among those suffering from schizophrenia, possibly indicating a lack of flexibility in functional community configuration. It is possible that heterogeneous methodological choices influence the discrepancies in modularity findings, specially given the difficulties to assess the quality of the partitions produced by any algorithm (regarding inconsistent results in reported modularity between early onset). 

As stated before, the analysis of the Rand Index computed over the partitions obtained with the Surprise algorithm reveals that healthy patients display a larger variability in terms of their community structure which can be considered more reconfigurable. At the same time, schizophrenia patients display a more homogeneous distribution of partitions among the network nodes implying a less reconfigurable and thus more rigid functional topology. This is in accordance with existing literature\cite{reinen2018human}.

The probability functions related to edge weights (i.e. correlation in activity between pairs of ROIs) differ between groups, with each one having a distinct peak. In particular, the peaks of the distribution had a higher value in chronic schizophrenia patients, compared to controls and those who had received a recent diagnosis. This could imply that connectivity between pairs of regions has the tendency progress with disease.

fMRI literature on schizophrenia often reports a lower overall connectivity in schizophrenia \cite{zhou2018altered}. This is likely given by the higher sparsity of patient derived networks. However, as highlighted above, the existing connections tend to display a stronger weight profile. Notably, in spite of the higher sparsity, patients display a higher robustness to link removal by means of edge weight thresholding. This aspect may be related to the networks structure of the healthy subjects which may display a relatively connected core together with a more uniform distribution of edge weights.

\section*{Future Work}

As a follow up to the present paper the following steps are proposed:
\begin{itemize}
\item Pursuing a resilience analysis based on static/dynamic failures of nodes/edges.
\item Computing the differences in efficiency of communication among brain areas by means of global/local efficiency measure.
\item Considering spatial location and anatomical characterization of the network nodes and of the subjects characteristics (i.e. add to the analysis further variables represented by the meta-data of both the nodes and the network).
\item The present study allows for a proper characterization of the synchronization distribution of both populations (HC and SZ). A follow step would be the extraction of both distributions, in order to allow for further simulations of resting state dynamics.
\item Examining whether the results found here are time dependent. Studying MEG/EEG could allow to answer this question, and these results should be recovered in slow frequency bands. 
\end{itemize}

\section*{Acknowledgements}

This work is the output of the Complexity72h workshop,
held at IMT School in Lucca, 7-11 May 2018. \url{https://complexity72h.weebly.com/}

\bibliography{Refs}


\end{document}